\documentclass[twocolumn,english,aps,prb]{revtex4-1}
\usepackage{color}
\usepackage{amsmath}
\usepackage{graphicx}
\usepackage{amssymb}

\makeatletter
\@ifundefined{textcolor}{}
{%
 \definecolor{BLACK}{gray}{0}
 \definecolor{WHITE}{gray}{1}
 \definecolor{RED}{rgb}{1,0,0}
 \definecolor{GREEN}{rgb}{0,1,0}
 \definecolor{BLUE}{rgb}{0,0,1}
 \definecolor{CYAN}{cmyk}{1,0,0,0}
 \definecolor{MAGENTA}{cmyk}{0,1,0,0}
 \definecolor{YELLOW}{cmyk}{0,0,1,0}
 }

\@ifundefined{definecolor}
 {\usepackage{color}}{}
\@ifundefined{definecolor}{\@ifundefined{definecolor}
 {\usepackage{color}}{}
}{}

\newcommand{\ba}{\begin{eqnarray*}}
\newcommand{\ea}{\end{eqnarray*}}
\newcommand{\baa}{\begin{eqnarray}}
\newcommand{\eaa}{\end{eqnarray}}
\newcommand{\bea}{\begin{eqnarray}}
\newcommand{\eea}{\end{eqnarray}}
\newcommand{\be}{\begin{equation}}
\newcommand{\ee}{\end{equation}}

\newcommand{\sm}{SmB$_6$}
\newcommand{\pu}{PuB$_6$}

\newcommand{\br}{\mathbf{r}}
\DeclareMathOperator{\Tr}{Tr}
\DeclareMathOperator{\I}{Im}

\makeatother

\usepackage{babel}

\begin{document}

\title{
Scanning tunneling spectroscopy and surface quasiparticle interference\\
in models for the strongly correlated topological insulators {\sm} and {\pu}
}

\author{Pier Paolo Baruselli}
\author{Matthias Vojta}
\affiliation{Institut f\"ur Theoretische Physik, Technische Universit\"at Dresden, 01062 Dresden, Germany}


\begin{abstract}
{\sm} is one of the candidate compounds for topological Kondo insulators, a class of materials which combines a non-trivial topological band structure with strong electronic correlations. Here we employ a multiband tight-binding description, supplemented by a slave-particle approach to account for strong interactions, to theoretically study the surface-state signatures in scanning tunneling spectroscopy (STS) and quasiparticle interference (QPI). We discuss the spin structure of the three surface Dirac cones of {\sm} and provide concrete predictions for the energy and momentum dependence of the resulting QPI signal. Our results also apply to {\pu}, a strongly correlated topological insulator with a very similar electronic structure.
\end{abstract}

\date{\today}


\maketitle


Topological insulators (TIs) with strong correlations are considered to be of crucial importance in the exciting field of topological phases: They may provide TI states which are truly bulk-insulating -- a property not easily realized in current Bi-based TIs -- and they may host novel and yet unexplored interaction-driven phenomena.

In this context, the material {\sm} has attracted enormous attention recently, as it has been proposed\cite{takimoto,lu_smb6_gutz, tki_cubic} to realize a three-dimensional (3D) topological Kondo insulator (TKI), i.e., a material where $f$-electron local moments form at intermediate temperatures $T$ and are subsequently screened at low $T$, such that a topologically non-trivial bandstructure emerges from Kondo screening.\cite{tki1}

While a number of experiments on {\sm}, such as transport studies,\cite{wolgast_smb6,fisk_smb6_topss, smb6_junction_prx} quantum oscillation measurements,\cite{lixiang_smb6} angle-resolved photoemission spectroscopy (ARPES),\cite{neupane_smb6,mesot_smb6, smb6_arpes_feng,smb6_arpes_reinert,smb6_past_allen} and scanning tunneling spectroscopy (STS)\cite{pnas_smb6_stm,hoffman_smb6} appear consistent with the presence of Dirac-like surface states expected in a TKI, a direct proof of their topological nature has been lacking until recently. Moreover, doubts have been raised about the proper interpretation of ARPES data.\cite{sawatzky_smb6,smb6_prx_arpes}

Two types of experiments are usually considered as smoking-gun probes of TI surface states: (i) spin-resolved ARPES which can detect the spin-momentum locking of the surface states\cite{Fu2007,tirev1,tirev2} and (ii) Fourier-transform STS (FT-STS) which can detect the absence of backscattering\cite{Fu2007} in quasiparticle interference (QPI) patterns which is a direct consequence of the spin-momentum locking.\cite{yazdani,kapitulnik,xue,xia}
Very recently, spin-resolved ARPES has successfully been applied to {\sm} and has confirmed spin-momentum locking of the surface states.\cite{smb6_arpes_mesot_spin}
In contrast, to date no high-quality FT-STS exist on {\sm} as well as on other candidate TKI materials, such as {\pu}.\cite{pub6}

It is the purpose of this paper to provide concrete predictions for FT-STS measurements on cubic TKIs. To this end we study the physics of local defects in a multiband Anderson lattice model for {\sm} and \pu, whose tight-binding (TB) part is derived from band-structure calculations.
We determine the spin structure of the three surface Dirac cones and discuss the momentum dependence of the resulting QPI signal for different types of scatterers. Our results may be directly tested in future FT-STS experiments on {\sm} and {\pu}.

%




\begin{figure}[tbp]
\includegraphics[width=0.38\textwidth]{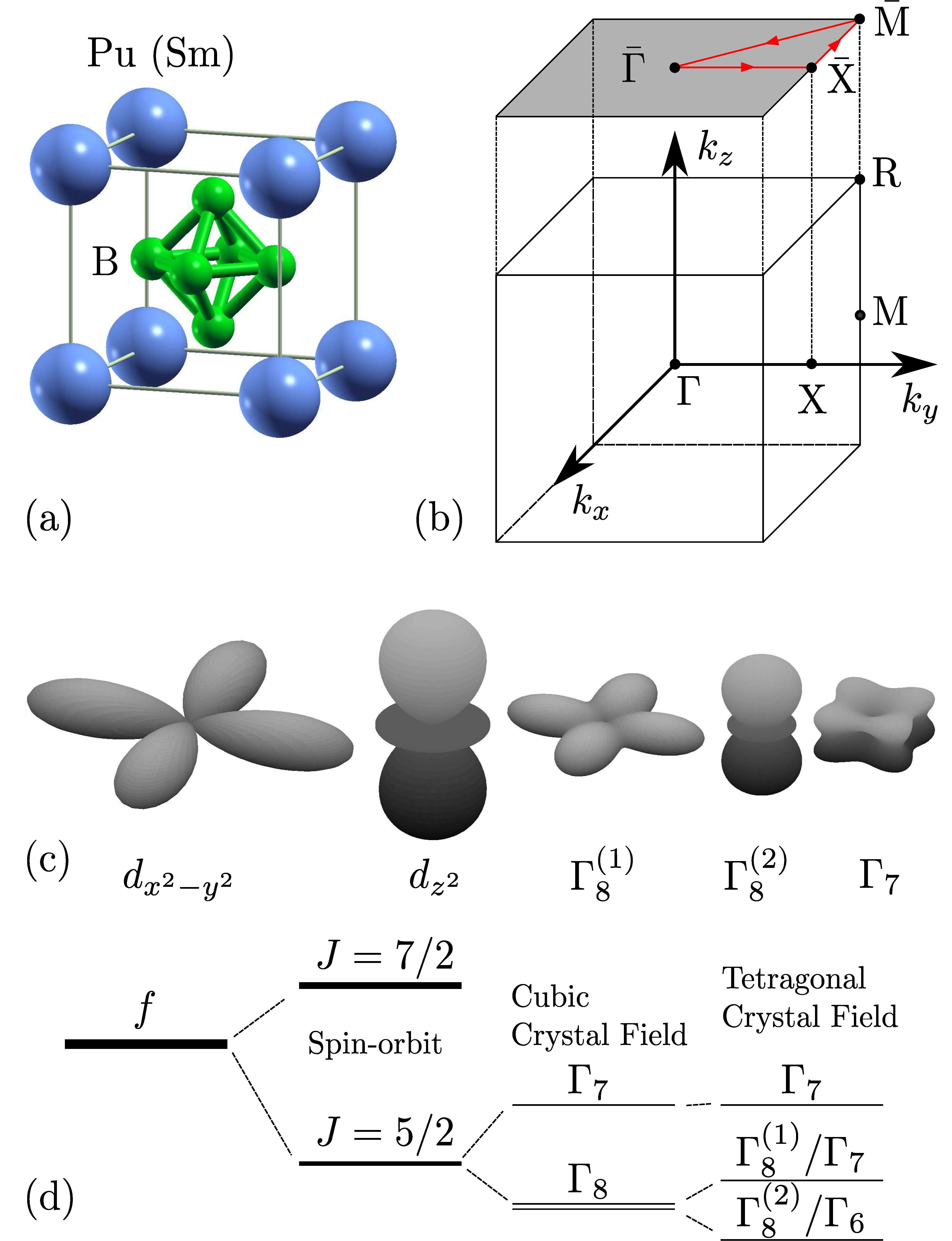}
\caption{
(a) Cubic crystal structure of {\sm} and \pu.
(b) 3D  Brillouin zone and its projection to a 2D Brillouin zone for a (001) surface.
(c) The 5 orbitals used in the TB model (all of them Kramers-degenerate).
(d) Schematic evolution of the $f$ levels under spin-orbit and crystal-field interactions. The tetragonal splitting is relevant near a surface.
}
\label{fig_struc}
\end{figure}

{\em Multi-orbital Anderson model.}
To describe the electronic properties of both {\sm} and \pu, which possess the same CsCl-like lattice structure, Fig. \ref{fig_struc}(a), and a very similar bandstructure, we employ a generalized version of the TB model of Refs. \onlinecite{takimoto} and \onlinecite{tki_cubic}.
The model entails a total of 10 rare-earth orbitals per site, namely the spin-degenerate $E_g$ ($d_{x^2-y^2}$ and $d_{z^2}$) quadruplet and the lowest-lying $f$-shell $J=5/2$ multiplet, see Fig. \ref{fig_struc}(c)-(d).
Other orbitals, including the rare-earth $J=7/2$ multiplet and all B$_6$ states, are excluded, since ab-initio methods show that their energies are far away from the Fermi level.\cite{lu_smb6_gutz, smb6_korea,pub6}
The cubic crystal field splits the $J=5/2$ multiplet into a $\Gamma_8$ quadruplet and a $\Gamma_7$ doublet, which read
$|\Gamma_8^{(1)}\pm\rangle = \sqrt\frac{5}{6}|\pm\frac{5}{2}\rangle + \sqrt\frac{1}{6}|\mp\frac{3}{2}\rangle$,
$|\Gamma_8^{(2)}\pm\rangle = |\pm\frac{1}{2}\rangle$,
$|\Gamma_7\pm\rangle = \sqrt\frac{1}{6}|\pm\frac{5}{2}\rangle -
\sqrt\frac{5}{6}|\mp\frac{3}{2}\rangle$
where $\pm$ denotes the pseudo-spin index.

The total Anderson Hamiltonian is:
\be\label{h_tot}
H_0=H_{dd}+H_{df}+H_{ff}+H_U
\ee
with $H_U$ encoding the local interaction and
\begin{align}
H_{dd}&=\sum_{i\sigma\alpha} \epsilon_{\alpha}^d  d^\dagger_{i\sigma\alpha}d_{i\sigma\alpha}- \!\! \sum_{\langle ij \rangle \sigma\alpha\alpha'} \!\!\! t^d_{ij\sigma\alpha\alpha'}(d_{i\sigma\alpha}^\dagger d_{j\sigma\alpha'}+h.c.), \notag\\
%
H_{ff}&=\sum_{i\sigma\alpha} \epsilon_{\alpha}^f f^\dagger_{i\sigma\alpha}f_{i\sigma\alpha}- \!\!\! \sum_{\langle ij \rangle \sigma\sigma'\alpha\alpha'}  \!\!\!\! (t^f_{ij\sigma\sigma'\alpha\alpha'}f_{i\sigma\alpha}^\dagger f_{j\sigma'\alpha'}+h.c.),\notag\\
\label{hdf}
H_{df}&=\sum_{\langle ij \rangle \sigma\sigma'\alpha\alpha'} \!(V_{ij\sigma\sigma'\alpha\alpha'} d_{i\sigma\alpha}^\dagger f_{j\sigma'\alpha'}+h.c.),
\end{align}
being the $d$ and $f$ kinetic energies and the hybridization, respectively.
Here, $\sigma$ and $\alpha$ denote the (pseudo)spin and orbital degrees of freedom, so in the $d$ shell $\sigma=\uparrow,\downarrow$ and $\alpha=d_{z^2}, d_{x^2-y^2}$, while in the $f$ shell $\sigma=+,-$ and $\alpha=\Gamma_8^{(1)}, \Gamma_8^{(2)}, \Gamma_7$.

Hopping and hybridization terms in $\langle ij\rangle$ are included up to 7th nearest neighbor (NN) sites, with $|\br_i-\br_j|\leq \sqrt{9}$. All parameter values were taken from the ab-initio calculations of Ref.~\onlinecite{pub6}, obtained by projecting LDA results to maximally localized Wannier functions. While these calculation are for {\pu}, our results should also apply to {\sm} -- perhaps with an adjustment of the overall energy scale, see below -- given the strong similarities of the two materials.\cite{pub6,haulepriv}
The concrete values of $t_{ij}$ and $V_{ij}$ up to 2nd NN are given in the supplement.\cite{suppl_mat}
%


{\em Hubbard repulsion and slave-boson approximation.}
The $f$ electrons are subject to a strong Coulomb repulsion $H_U$. Here we employ the standard slave-boson approximation which implements reduced charge fluctuations in the infinite-repulsion limit at the mean-field level.\cite{read_newns_slavebosons_1,read_newns_slavebosons, hewson}
For both {\sm} and {\pu} the dominant charge configurations are $d^1 f^5$ and $d^0 f^6$, such that it is convenient to work in a hole representation: The Coulomb repulsion suppresses states with more than one $f$ hole per site. The remaining states of the local $f$ Hilbert space are represented by auxiliary particles, $b_i$ and $\tilde{f}_{i\sigma\alpha}$ for $f^6$ and $f^5$ states, respectively. At the mean-field level, $b_i \rightarrow b=\langle b_i \rangle$ is condensed, and a Lagrange multiplier $\lambda$ is used to impose the required Hilbert-space constraint. Both parameters need to be determined self-consistently, together with the overall chemical potential; technical details can be found in the supplement.\cite{suppl_mat}

This procedure transforms the Anderson model of Eq.~\eqref{h_tot} into a non-interacting TB model, with the influence of the Coulomb repulsion encoded in a downward renormalization of the $f$ kinetic energy by a factor $b^2$ and the hybridization by a factor $b$. In addition, the $f$-level energy $\epsilon^f_{\alpha}$ is shifted towards the Fermi level.


{\em STS, Defects, and QPI.}
To calculate the STS signal on a (001) surface, we solve the renormalized TB model in a slab geometry. We ignore a possible surface reconstruction, but comment on its effects below (note that the unreconstructed (001) surface of {\sm} is polar\cite{sawatzky_smb6}).

In order to model QPI, we take into account scattering off isolated defects which we assume to be located in the surface layer. For simplicity, we take point-like scatterers and neglect the local modifications of the slave-boson parameters.\cite{prb_io_tki} Impurity-induced changes of electron propagators are calculated using a T-matrix formalism, with details given in the supplement.\cite{suppl_mat}

The output quantity is the Green's function ${G}_{aa'}(E,\br,\br')$, which depends on the energy $E$, on the positions $\br$ and $\br'$, and on the orbital indices $a,a'=1,\ldots,10$.
The local density of states (LDOS) is the (orbital) trace of the imaginary part of the local Green's function, $\rho(E,\br)=-1/\pi\I \Tr \hat{G}(E,\br,\br)$.
However, the STS signal is not simply proportional to the LDOS, as the tip samples each orbital with a different weight, and interference effects are also present.\cite{madhavan98,kroha00,coleman09,morr10,balatsky_kondohole}
To simulate this process, in the spirit of the \textit{cotunneling} of Ref.~\onlinecite{coleman09}, we compute
$\rho_{STS}(E,\br)=-1/\pi \I\Tr [\hat\psi \hat G(E,\br,\br)  \hat\psi^T]$,
where $\hat\psi$ is a $4\times 10$ matrix containing the coupling between each of the 10 orbitals to each of four assumed tip-electron channels (two spin directions and two orbitals); for details see supplement.\cite{suppl_mat}
%
The QPI signal $\rho_{QPI}(E,k_x,k_y,z\!=\!1)$ is then obtained from $\rho_{STS}(E,x,y,z\!=\!1)$ by a Fourier transform in the $xy$ plane; $\rho_{QPI}$ is real for the single-impurity case considered here.


\begin{figure}[tbp]
\includegraphics[width=0.43\textwidth]{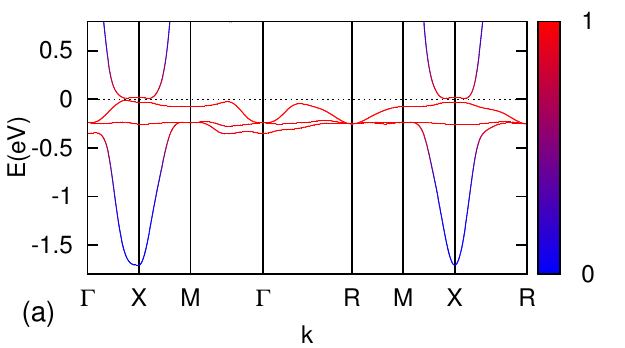}
\includegraphics[width=0.41\textwidth]{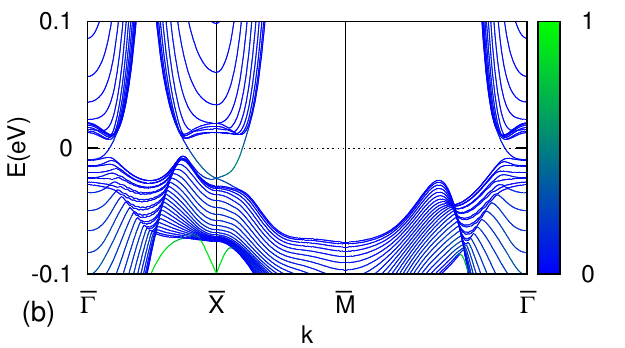}
\caption{
(a) Bulk dispersion from the renormalized TB model along a path in the 3D Brillouin zone; the color code shows the $f$ weight.
(b) Band structure of a $n_z=25$ slab along a path in the 2D (surface) Brillouin zone; the color code shows the spectral weight in the topmost ($z\!=\!1$) layer.
}
\label{fig_bands}
\end{figure}

{\em Results: Band structure and surface states.}
Fig. \ref{fig_bands}(a) shows the 3D bandstructure as obtained from the renormalized TB model. The $d$ band has a minimum at about $-1.7$~eV at the X point, as observed in ARPES experiments for {\sm},\cite{neupane_smb6,mesot_smb6, smb6_arpes_feng,smb6_arpes_reinert,smb6_past_allen}
while $f$ states lie close to the Fermi energy.
Around the $X$ point, the bottom of the conduction (top of the valence) band is mainly of $\Gamma_7$ ($\Gamma_8$) character.
The overall agreement with DFT calculations, possibly with many-body corrections, \cite{lu_smb6_gutz,pub6,smb6_korea} is satisfactory,
even though reproducing some finer details
would require including even longer-range hoppings; we have verified that this does not significantly alter surface states and QPI spectra.
We note that, according to DMFT calculations,\cite{pub6} the interaction-induced renormalization factor of the $f$ kinetic energy should be $\sim 0.2$ rather than our $b^2 \sim 0.5$. Furthermore, LDA results indicate that that $f$-band energies are by a factor of $1.5\ldots2$ smaller in {\sm} as compared to {\pu}.\cite{lu_smb6_gutz,pub6} As a consequence, a rescaling of the bulk energies close to the Fermi level by a factor $\sim 0.2 \ldots 0.4$
might be necessary for a quantitative comparison with {\sm} experiments. We stress, however, that this does not strongly affect the {\em momentum} dependence of the QPI spectra to be discussed below.

By computing topological indices\cite{tki2,Fu2007,fu_kane_ti_invsym} it is easy to show that the renormalized TB model is a strong topological insulator for the range of parameters pertinent to {\sm} \cite{lu_smb6_gutz, takimoto,tki_cubic} and \pu.\cite{pub6}
Band inversion between even $d$ and odd $f$ bands occurs at the three inequivalent X points. As a result, three surface Dirac cones appear at the two $\bar{X}$ points and at $\bar\Gamma$ of the 2D surface Brillouin zone,\cite{lu_smb6_gutz,takimoto,tki_cubic,pub6} see Fig.~\ref{fig_bands}(b).
We obtain the Dirac energies to be $\epsilon_{\bar{\Gamma}}=-9$~meV and $\epsilon_{\bar{X}}=-24$~meV
and the Fermi momenta $k_{F\bar{\Gamma}}=0.15$~\AA$^{-1}$  and $k_{F\bar{X}}=0.19 - 0.17 $~\AA$^{-1}$ (we have used the {\sm} lattice constant 4.13\AA).
Experimental results from ARPES for {\sm} are\cite{smb6_arpes_feng,neupane_smb6,smb6_past_allen}
$\epsilon_{\bar{\Gamma}}=-23$~meV, $\epsilon_{\bar{X}}=-65$~meV,
$k_{F\bar{\Gamma}}=0.09$~\AA$^{-1}$, $k_{F\bar{X}}=0.39-0.28$~\AA$^{-1}$.
While this agreement does not appear perfect, we note that the experimental estimates for $\epsilon_{\bar{\Gamma}}$  and $\epsilon_{\bar{X}}$ were obtained by a linear extrapolation of the low-$E$ dispersion;\cite{smb6_arpes_feng} the curvature in our surface bands indicates that this might be unwarranted. In addition, the precise dispersion of surface states sensitively depends on many factors which are difficult to take into account in a microscopic model. These include modified orbital energies, a modified crystal field, and modified Kondo screening\cite{prb_io_tki} near the surface as well as surface termination, surface reconstruction, and disorder.
In particular, the unreconstructed (001) surface of {\sm} is polar, showing also surface states of non-topological origin, while the $2\times 1$ reconstructed surface is non-polar,\cite{hoffman_smb6, pnas_smb6_stm} and is the one which more closely resembles our modelling (ignoring reconstruction effects such as band-folding \cite{mesot_smb6}).
The dependence of in-gap states on the surface termination has also been noted in ab-initio calculations.\cite{kimkim14}

\begin{figure}[tbp]
\includegraphics[width=0.48\textwidth]{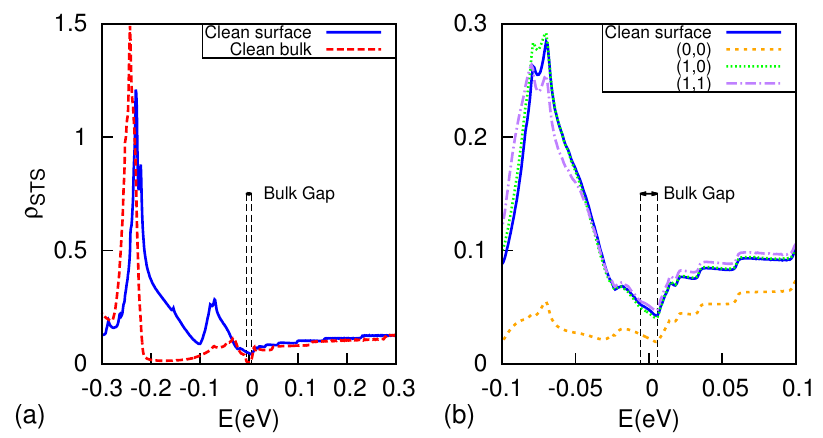}
\caption{(a) STS signal of a clean surface, together with the hypothetical signal from the bulk obtained by using periodic boundary conditions along $z$.
The residual bulk signal inside the gap, between $-6$ and $6$~meV, arises from the finite Lorentzian broadening used in the calculation. The bulk signal's large peak around $-0.3$~eV originates from weakly dispersing $f$ states, Fig.~\ref{fig_bands}(a).
(b) STS signal over a Kondo hole located at $(0,0)$ and in its proximity at $(1,0)$ and $(1,1)$, compared to the signal of the clean surface as in (a).
}
\label{fig_stm}
\end{figure}


{\em Results: STS signal.}
The energy-dependent STS signal, Fig.~\ref{fig_stm}(a), shows a pseudogap close to the Fermi energy; at negative (positive) energies the signal originates mainly from $f$ ($d$) states.
Existing STS experiments on $2\times 1$ reconstructed (001) {\sm} surfaces\cite{hoffman_smb6, pnas_smb6_stm} show a peak at roughly $-8$~meV and a dip near $E_F$, leading to a Fano-like structure.
Its shape and peak-to-background ratio are very similar similar to that in our calculation.
However, our peak lies considerably deeper in energy, at about $-80$~meV, corresponding a set of surface states, while at about $-30$~meV, where bulk $f$ states show an LDOS peak, we see no peak in the surface signal.
As noted above, surface states are extremely sensitive to the local environment, and changes in their dispersion will strongly influence the STS signal:
For example, the unreconstructed (polar) surface of {\sm} displays a peak at $-28$~meV (instead of $-8$~meV), and disordered surfaces show even more complex behavior.\cite{hoffman_smb6}
As a consequence, we believe the peaks observed in experiments at $-8$~meV or $-28$~meV arise from surface (rather than bulk) states, and apparently require a more accurate modelling of states far from the Dirac points.

Near a Kondo hole, i.e., a defect with missing $f$ orbital, the tunneling spectrum is mainly suppressed at negative energies where the signal has $f$ character, Fig.~\ref{fig_stm}(b). No resonance peaks occur for these strong scatterers, due to the large particle--hole asymmetry of the $f$ band.\cite{prb_io_tki} We note that low-energy resonances may still occur for scatterers of fine-tuned intermediate strength.



\begin{figure}[tbp]
\includegraphics[width=0.49\textwidth]{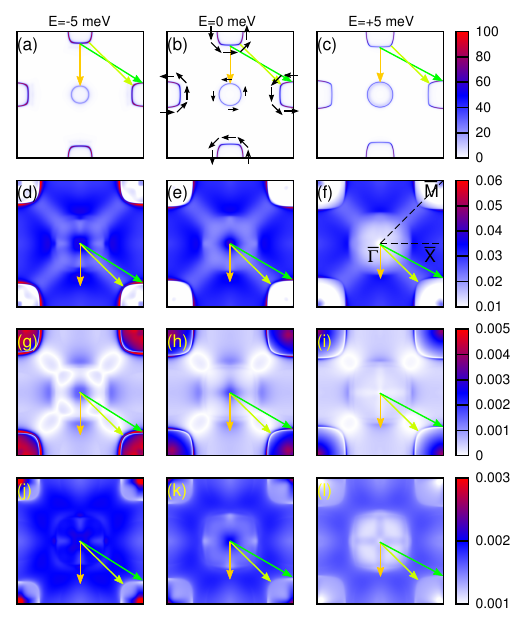}
\caption{(a-c) ARPES signal at $-5$~meV, $0$, and $+5$~meV,
and corresponding QPI signal $|\rho_{QPI}|$ for
(d-f) a Kondo hole,
(g-i) a weak $\Gamma_7$ scatterer, and
(j-l) a weak $\Gamma_8$ scatterer, both in the Born approximation with $V=10$~meV. 
In panel (b) we schematically show the surface-state spin structure,\cite{suppl_mat}
which agrees with the experimental results of Ref.~\onlinecite{smb6_arpes_mesot_spin}.
}
\label{fig_qpi}
\end{figure}

{\em Results: QPI signal.}
In Fig.~\ref{fig_qpi} we show the QPI signal inside the bulk gap for different types of impurities, with the corresponding surface ARPES signal for comparison; the figure also indicates the spin polarization of the surface states.\cite{suppl_mat} Notably this spin structure agrees with the recent results of spin-resolved ARPES on {\sm}.\cite{smb6_arpes_mesot_spin}

As is common for all TIs, the QPI signal from intracone scattering due to non-magnetic impurities is weak and non-peaked near the Dirac point,\cite{tirev1,tirev2,guo_franz_sti} as backscattering $\bf k \leftrightarrow -\bf k$  involves states with opposite spin. No such argument holds for intercone scattering, which, consequently, can give rise to pronounced peaks in the QPI signal.\cite{prb_io_tki}
Remarkably, the Dirac-cone spin structure of Fig. \ref{fig_qpi}(b) is such that {\em also} intercone scattering tends to be suppressed:
This is because the spin directions for pairs of stationary points\cite{liu_qpi_spa}
(i.e. points with parallel tangents to their constant-energy contour, e.g., the ones connected by colored arrows in Fig.~\ref{fig_qpi}(a,b,c)),
are essentially antiparallel, which applies both to $\bar\Gamma$--$\bar X$ and $\bar X$--$\bar X'$ scattering.\cite{comment_M}
However, care is required: While QPI spectra associated to $\Gamma_8$ impurities are mostly non-peaked, Fig.~\ref{fig_qpi}(j,k,l), as suggested by the spin structure,
both $\Gamma_7$ impurities and Kondo holes do give rise to QPI peaks corresponding to $\bar X$--$\bar X'$ scattering, see Fig.~\ref{fig_qpi}(d-i).
As shown in the supplement,\cite{suppl_mat} this can be ascribed to the $\Gamma_7$ component of the surface states which in fact displays parallel spin expectation values at certain pairs of stationary points, allowing for efficient $\bar X$--$\bar X'$ scattering.
Thus, details of the intercone signal depend on the character of the scattering center, which might help to experimentally identify different scatterers.

The energy dependence of the QPI signal within the bulk gap is weak: Upon increasing the energy, the intercone scattering momenta shrink, and the overall signal strength decreases.
Upon leaving the bulk gap, we expect a rapid decrease of the surface QPI signal, due to the hybridization of surface with bulk states.

To underline how sensitively the QPI signal depends on proper modelling, in particular on the Dirac-cone spin structure, we have repeated the same calculation with models of reduced $f$-orbital content, i.e., retaining only the $\Gamma_7$ doublet or only the $\Gamma_8$ quartet in the model Hamiltonian \eqref{h_tot}, as in Fig.~5 of Ref.~\onlinecite{pub6}. The resulting QPI patterns drastically differ, and the ``only $\Gamma_7$'' case even yields a spin structure in disagreement with experiment.\cite{smb6_arpes_mesot_spin} Details are in the supplement.\cite{suppl_mat}

These results show that the orbital content of both surface states and impurities are relevant to QPI spectra. Importantly, this cannot be properly captured in effective low-energy models.
In particular, the relative $\Gamma_7$/$\Gamma_8$ weight of both the Dirac-cone states and the impurities determine the strength of the $\bar X$--$\bar X'$ scattering peak of experimental QPI spectra.
We note that none of the calculations showed a significant QPI signal for scattering between the $\bar \Gamma$ and $\bar X$ cones.


{\em Summary.}
We have computed ARPES, STS, and QPI spectra within a renormalized multiorbital TB model for the strongly-correlated TI materials {\sm} and {\pu}.
Both ARPES and STS spectra agree semi-quantitatively with existing experimental results for {\sm}.
The remaining disagreement can be attributed to modelling uncertainties concerning the interaction-induced renormalization of the kinetic energy and the detailed electronic structure of the surface,
where surface termination and reconstruction play an important role.

We have made concrete predictions for the QPI signal. We have found that QPI peaks corresponding to $\bar X$--$\bar X'$ intercone scattering can appear for particular types of impurities, which can be related to the spin structure and orbital content of the Dirac-cone states.
We have also considered a model variant which results in a spin structure in disagreement with experiment\cite{smb6_arpes_mesot_spin} and yields a qualitatively different QPI signal, illustrating that QPI is a powerful probe for the surface spin structure of TIs with multiple Dirac cones.
Hence, the observation of a weakly peaked low-energy QPI signal in {\sm}, possibly with $\bar X$--$\bar X'$ scattering peaks, would not only confirm the topological nature of the surface states, but also the Dirac-cone spin structure as reported in Ref.~\onlinecite{smb6_arpes_mesot_spin}.

Future work should include a more detailed modelling of surface effects as well as a study of finite-temperature crossovers, similar to Refs.~\onlinecite{benlagra11,assaad_temperature}.



We thank H. Fehske, L. Fritz, and D. K. Morr for discussions and collaborations on related work. Furthermore, we are very grateful to X. Deng, K. Haule, and G. Kotliar for sharing their data on \pu.
This research was supported by the DFG through FOR 960 and GRK 1621
as well as by the Helmholtz association through VI-521.

\bibliographystyle{apsrev4-1}
\bibliography{tki}

\begin{thebibliography}{46}
\bibitem [{com()}]{comment_M}%
  \BibitemOpen
  \href@noop {} {}\bibinfo {note} {Intercone $\bar X$--$\bar X'$ scattering
  does not lead to a strong QPI peak near $(\pi,\pi)$, despite the almost
  perfect nesting of the two $\bar{X}$ cones: For isotropic cones, intercone
  scattering by $(\pi,\pi)$ is equivalent to intracone scattering by $(0,0)$,
  and the latter produces\cite{guo_franz_sti} a flat signal near $(0,0)$
  only.}\BibitemShut {Stop}%
\end{thebibliography}

\end{document}


\title{
Supplemental material for:\\
Scanning tunneling spectroscopy and surface quasiparticle interference\\
in models for the strongly correlated topological insulators {\sm} and {\pu}
}

\author{Pier Paolo Baruselli}
\author{Matthias Vojta}
\affiliation{Institut f\"ur Theoretische Physik, Technische Universit\"at Dresden, 01062 Dresden, Germany}

\date{\today}

\maketitle

\section{Tight-binding parameters}

Our tight-binding (TB) model, Eq.~{\eqtb} of the main text, includes hopping and hybridization terms up to 7th nearest-neighbor (NN) sites of the cubic lattice, i.e., up to cartesian distances $(300)$ and $(221)$.
We note that 1st and 2nd NN terms are needed to yield the minimum of the $d$ band and the maximum of the $\Gamma_8$ band to be located both at $X$, while the 3rd NN is needed for a proper description of the $\Gamma_7$ band.
Further NN terms are needed for a quantitative adjustment of the low-energy bandstructure.

In the following we sketch the construction of the model and specify its parameters for first and second-neighbor terms. The numerical values for all parameters were taken from tight-binding fits to the ab-initio results for {\pu} of Ref.~\onlinecite{pub6}.

\subsection{On-site energies}

We start with the local (i.e. on-site) orbital energies which enter $H_{dd}$ and $H_{ff}$ in Eq.~{\eqtb} of the main text. Their values are $\epsilon_{\Gamma_8}^f\equiv\epsilon_{\Gamma_8^{(1)}}^f=\epsilon_{\Gamma_8^{(2)}}^f=0.50$~eV $ \ne \epsilon_{\Gamma_7}^f=0.58$~eV, and $\epsilon_d\equiv\epsilon_{z^2}^d=\epsilon_{x^2-y^2}^d=2.47$~eV.

Near a surface, the crystal-field symmetry is reduced, such that $\epsilon_{\Gamma_8^{(1)}}^f\ne\epsilon_{\Gamma_8^{(2)}}^f$ and $\epsilon_{z^2}^d\ne \epsilon_{x^2-y^2}^d$ is expected. Considering the lack of corresponding ab-initio results, we have ignored this effect, but we note that the hopping and hybridization terms in our TB model effectively generate such a surface-induced splitting.

\subsection{First NN}

The nearest-neighbor processes connect sites with cartesian distances $(\pm1,0,0)$, $(0,\pm1,0)$, $(0,0,\pm1)$.
In what follows we abbreviate $\cos k_x\equiv c_x$, $\cos k_y\equiv c_y$, $\cos k_z\equiv c_z$, $\sin k_x\equiv s_x$, $\sin k_y\equiv s_y$, $\sin k_z\equiv s_z$.
Moreover, to shorten notation, we specify energies using the following ``units'': $\td=1$~eV, $\tf=0.01$~eV, $\vv=0.1$~eV.

To efficiently generate the hopping piece for the cubic-symmetry case, we follow the treatment of Ref.~\onlinecite{tki_cubic}. We fix the matrix elements along the $(001)$ direction, then apply a rotation in orbital space according to
\be\label{uxd}
U_x^d=\frac{e^{-i\pi/4}}{2\sqrt{2}}
\left( \begin{array}{llll}
-1 & i &\sqrt{3} &-i\sqrt{3} \\
-1 & -i &\sqrt{3} &i\sqrt{3} \\
-\sqrt{3} &i\sqrt{3}& -1 & i\\
-\sqrt{3} &-i\sqrt{3}& -1 & -i
\end{array}\right)
\ee
and
\be\label{uxf}
U_x^f=\frac{e^{-i\pi/4}}{2\sqrt{2}}
\left( \begin{array}{llllll}
-1 & i &\sqrt{3} &-i\sqrt{3}&0&0 \\
-1 & -i &\sqrt{3} &i\sqrt{3}&0&0 \\
-\sqrt{3} &i\sqrt{3}& -1 & i&0&0\\
-\sqrt{3} &-i\sqrt{3}& -1 & -i&0&0\\
0&0&0&0&2& -2i\\
0&0&0&0&2& 2i
\end{array}\right)
\ee
to obtain the matrix elements in the $(100)$ direction, and finally use $U_y^d=U_x^d\cdot U_x^d$, $U_y^f=U_x^f\cdot U_x^f$ for the $(010)$ direction.

\begin{widetext}
The resulting $H_{dd}^{1}$ is diagonal in spin space and reads in the $d_{x^2-y^2}$, $d_{z^2}$ basis:
\bea
H_{dd}^{1}=-\td
\left( \begin{array}{ll}
(c_x+c_y)(\frac{1}{2}\eta_{x}^{d1}+\frac{3}{2}\eta_{z}^{d1})+2c_z\eta_{x}^{d1}& \frac{\sqrt{3}}{2}(c_x-c_y)(\eta_{x}^{d1}-\eta_{z}^{d1})\\
\frac{\sqrt{3}}{2}(c_x-c_y)(\eta_{x}^{d1}-\eta_{z}^{d1})& (c_x+c_y)(\frac{1}{2}\eta_{z}^{d1}+\frac{3}{2}\eta_{x}^{d1})+2c_z\eta_{z}^{d1}
\end{array}\right).
\eea
Here $\eta_x^{d1}=-0.089$, $\eta_z^{d1}=0.807$ are the numerical hopping parameters extracted from Ref.~\onlinecite{pub6} in units of $\td$.

Similarly $H_{ff}^{1}$ is diagonal in pseudospin space, and reads in the $\Gamma_{8}^{{(1)}}$, $\Gamma_{8}^{{(2)}}$, $\Gamma_{7}$ basis:
\bea
H_{ff}^{1}=-\tf
\left( \begin{array}{lll}
(c_x+c_y)(\frac{1}{2}\eta_{x}^{f1}+\frac{3}{2}\eta_{z}^{f1})+2c_z\eta_{x}^{f1}& \frac{\sqrt{3}}{2}(c_x-c_y)(\eta_{x}^{f1}-\eta_{z}^{f1})&-\eta_{78}^{f1}(c_x+c_y-2c_z)\\
\frac{\sqrt{3}}{2}(c_x-c_y)(\eta_{x}^{f1}-\eta_{z}^{f1})& (c_x+c_y)(\frac{1}{2}\eta_{z}^{f1}+\frac{3}{2}\eta_{x}^{f1})+2c_z\eta_{z}^{f1}&\sqrt{3}\eta_{78}^{f1}(-c_x+c_y)\\	
-\eta_{78}^{f1}(c_x+c_y-2c_z)& \sqrt{3}\eta_{78}^{f1}(-c_x+c_y)& 2\eta^{f1}_7 (c_x+c_y+c_z)
\end{array}\right)
\eea
with $\eta_{x}^{f1}=1.25$, $\eta_{z}^{f1}=-4.17$, $\eta^{f1}_7=-0.14$, $\eta^{f1}_{78}=-0.59$.

The hybridization $H_{df}^{1}$ is non-diagonal in spin space and reads
\be
V_{df}^{1}=i \vv
\left(
\begin{smallmatrix}
 2 \eta^{v1}_x s_z & \frac{1}{2} (\eta^{v1}_x+3 \eta^{v1}_z) (s_x-i s_y) & 0 & \frac{1}{2} \sqrt{3} (\eta^{v1}_x-\eta^{v1}_z) (s_x+i s_y) & 2 \eta^{v1}_7 s_z & -\eta^{v1}_7
   (s_x-i s_y) \\
 \frac{1}{2} (\eta^{v1}_x+3 \eta^{v1}_z) (s_x+i s_y) & -2 \eta^{v1}_x s_z & \frac{1}{2} \sqrt{3} (\eta^{v1}_x-\eta^{v1}_z) (s_x-i s_y) & 0 & -\eta^{v1}_7 (s_x+i s_y) & -2
   \eta^{v1}_7 s_z \\
 0 & \frac{1}{2} \sqrt{3} (\eta^{v1}_x-\eta^{v1}_z) (s_x+i s_y) & 2 \eta^{v1}_z s_z & \frac{1}{2} (3 \eta^{v1}_x+\eta^{v1}_z) (s_x-i s_y) & 0 & -\sqrt{3} \eta^{v1}_7 (s_x+i
   s_y) \\
 \frac{1}{2} \sqrt{3} (\eta^{v1}_x-\eta^{v1}_z) (s_x-i s_y) & 0 & \frac{1}{2} (3 \eta^{v1}_x+\eta^{v1}_z) (s_x+i s_y) & -2 \eta^{v1}_z s_z & -\sqrt{3} \eta^{v1}_7 (s_x-i s_y) &
   0
\end{smallmatrix}
\right)
\ee
where the basis is $\Gamma_{8}^{{(1)}}+$, $\Gamma_{8}^{{(1)}}-$, $\Gamma_{8}^{{(2)}}+$, $\Gamma_{8}^{{(2)}}-$, $\Gamma_{7}+$, $\Gamma_{7}-$ for columns, and
$d_{x^2-y^2}\uparrow$, $d_{x^2-y^2}\downarrow$, $d_{z^2}\uparrow$, $d_{z^2}\downarrow$ for rows.
The numerical hybridization parameters are $\eta^{v1}_x=0.422$, $\eta^{v1}_z=-2.11$, $\eta^{v1}_7=-0.166$.

\subsection{Second NN}

Second NN processes correspond to distances $(0,\pm1,\pm1)$, $(\pm1,0,\pm1)$, $(\pm1,\pm1,0)$.
In analogy to the above, we start with matrix elements along the $(110)$ direction, then rotate by matrices $U_x^d$, $U_x^f$ for the $(011)$ direction, and by matrices $U_y^d$, $U_y^f$ for the $(101)$ direction.

Among the resulting Hamiltonian pieces, only $H_{dd}^{2}$ is diagonal in spin space. The final matrices read
\bea
H_{dd}^{2}=-\td
\left(
\begin{array}{cc}
 (4 c_x c_y+(c_x+c_y) c_z) \eta^{d2}_x+3 (c_x+c_y) c_z \eta^{d2}_z &  -\sqrt{3} (c_x-c_y) c_z (\eta^{d2}_x-\eta^{d2}_z)\\
 -\sqrt{3} (c_x-c_y) c_z (\eta^{d2}_x-\eta^{d2}_z) & 3 (c_x+c_y) c_z \eta^{d2}_x+(4 c_x c_y+(c_x+c_y) c_z) \eta^{d2}_z  \\
\end{array}
\right)
\eea
with $\eta^{d2}_x=0.136$, $\eta^{d2}_z=-0.29$;
\bea
H_{ff}^{2}=-\tf
\left(
\begin{smallmatrix}\nonumber
 4 \eta^{f2}_x c_x c_y+(\eta^{f2}_x+3 \eta^{f2}_z) (c_x+c_y) c_z & 0 & -\sqrt{3} (\eta^{f2}_x-\eta^{f2}_z) (c_x-c_y) c_z-4 i \eta^{f2}_{xz} s_x s_y &\dots\\
 0 & 4 \eta^{f2}_x c_x c_y+(\eta^{f2}_x+3 \eta^{f2}_z) (c_x+c_y) c_z & 4 \eta^{f2}_{xz} (s_x-i s_y) s_z &\dots\\
 4 i \eta^{f2}_{xz} s_x s_y-\sqrt{3} (\eta^{f2}_x-\eta^{f2}_z) (c_x-c_y) c_z & 4 \eta^{f2}_{xz} (s_x+i s_y) s_z  & 4 \eta^{f2}_z c_x c_y+(3 \eta^{f2}_x+\eta^{f2}_z) (c_x+c_y) c_z &\dots\\
 -4 \eta^{f2}_{xz} (s_x-i s_y) s_z & -\sqrt{3} (\eta^{f2}_x-\eta^{f2}_z) (c_x-c_y) c_z-4 i \eta^{f2}_{xz} s_x s_y & 0 &\dots\\
 4 \eta^{f2}_{x7} c_x c_y-2 \eta^{f2}_{x7} (c_x+c_y) c_z & 2 \sqrt{3} \eta^{f2}_{z7} (s_x-i s_y) s_z & 2 \sqrt{3} \eta^{f2}_{x7} (c_x-c_y) c_z-4 i \eta^{f2}_{z7} s_x s_y &\dots\\
 -2 \sqrt{3} \eta^{f2}_{z7} (s_x+i s_y) s_z & 4 \eta^{f2}_{x7} c_x c_y-2 \eta^{f2}_{x7} (c_x+c_y) c_z & -2 \eta^{f2}_{z7} (s_x-i s_y) s_z & \dots
\end{smallmatrix}
\right.
\eea
\bea
\left.
\begin{smallmatrix}
\dots&-4 \eta^{f2}_{xz} (s_x+i s_y) s_z & 4 \eta^{f2}_{x7} c_x c_y-2 \eta^{f2}_{x7} (c_x+c_y) c_z & -2 \sqrt{3} \eta^{f2}_{z7} (s_x-i s_y) s_z \\
\dots&4 i  \eta^{f2}_{xz} s_x s_y-\sqrt{3} (\eta^{f2}_x-\eta^{f2}_z) (c_x-c_y) c_z & 2 \sqrt{3} \eta^{f2}_{z7} (s_x+i s_y) s_z & 4 \eta^{f2}_{x7} c_x c_y-2 \eta^{f2}_{x7} (c_x+c_y) c_z \\
\dots&0 & 2 \sqrt{3} \eta^{f2}_{x7} (c_x-c_y) c_z+4 i \eta^{f2}_{z7} s_x s_y & -2 \eta^{f2}_{z7} (s_x+i s_y) s_z \\
\dots&4 \eta^{f2}_z c_x c_y+(3 \eta^{f2}_x+\eta^{f2}_z) (c_x+c_y) c_z & 2 \eta^{f2}_{z7} (s_x-i s_y) s_z & 2 \sqrt{3} \eta^{f2}_{x7} (c_x-c_y) c_z-4 i \eta^{f2}_{z7} s_x s_y \\
\dots&2 \eta^{f2}_{z7} (s_x+i s_y) s_z & 4 \eta^{f2}_{7} (c_y c_z+c_x (c_y+c_z)) & 0 \\
\dots&2 \sqrt{3} \eta^{f2}_{x7} (c_x-c_y) c_z+4 i \eta^{f2}_{z7} s_x s_y & 0 & 4  \eta^{f2}_{7} (c_y c_z+c_x (c_y+c_z))
\end{smallmatrix}
\right)
\eea
with $\eta^{f2}_x=-1.03$, $\eta^{f2}_z=2.25$, $\eta^{f2}_{xz}=0.55$, $\eta^{f2}_{x7}=-0.82$, $\eta^{f2}_{z7}=2.89$, $\eta^{f2}_{7}=2.46$; and
\bea
V_{df}^{2}=i \vv
\left(
\begin{smallmatrix}\nonumber
 \left(\eta^{v2}_{xx}+\sqrt{3} \eta^{v2}_{xz}+\sqrt{3} \eta^{v2}_{zx}+3 \eta^{v2}_{zz}\right) (c_x+c_y) s_z & 4 \eta^{v2}_{xx} (c_y s_x-ic_x s_y)+\left(\eta^{v2}_{xx}-\sqrt{3} \eta^{v2}_{xz}-\sqrt{3} \eta^{v2}_{zx}+3 \eta^{v2}_{zz}\right) c_z (s_x-i s_y)  &\dots\\
 4 \eta^{v2}_{xx} (c_y s_x+i  c_x s_y)+\left(\eta^{v2}_{xx}-\sqrt{3} \eta^{v2}_{xz}-\sqrt{3} \eta^{v2}_{zx}+3 \eta^{v2}_{zz}\right) c_z (s_x+i s_y) & -\left(\eta^{v2}_{xx}+\sqrt{3} \eta^{v2}_{xz}+\sqrt{3} \eta^{v2}_{zx}+3 \eta^{v2}_{zz}\right) (c_x+c_y) s_z &\dots\\
 -\left(\sqrt{3} \eta^{v2}_{xx}+3 \eta^{v2}_{xz}-\eta^{v2}_{zx}-\sqrt{3} \eta^{v2}_{zz}\right) (c_x-c_y) s_z & 4 \eta^{v2}_{zx} (c_y s_x+i c_x s_y)-\left(\sqrt{3} \eta^{v2}_{xx}-3 \eta^{v2}_{xz}+\eta^{v2}_{zx}-\sqrt{3} \eta^{v2}_{zz}\right) c_z (s_x+i s_y)&\dots\\
 4 \eta^{v2}_{zx} (c_y s_x-i c_x s_y)-\left(\sqrt{3} \eta^{v2}_{xx}-3 \eta^{v2}_{xz}+\eta^{v2}_{zx}-\sqrt{3} \eta^{v2}_{zz}\right) c_z (s_x-i s_y) & \left(\sqrt{3} \eta^{v2}_{xx}+3 \eta^{v2}_{xz}-\eta^{v2}_{zx}-\sqrt{3} \eta^{v2}_{zz}\right) (c_x-c_y) s_z &\dots
\end{smallmatrix}
\right.
\eea
\bea
\begin{smallmatrix}\nonumber
\dots&-\left(\sqrt{3} \eta^{v2}_{xx}-\eta^{v2}_{xz}+3 \eta^{v2}_{zx}-\sqrt{3} \eta^{v2}_{zz}\right) (c_x-c_y) s_z & 4 \eta^{v2}_{xz} (c_y s_x+i c_xs_y)-\left(\sqrt{3} \eta^{v2}_{xx}+\eta^{v2}_{xz}-3 \eta^{v2}_{zx}-\sqrt{3} \eta^{v2}_{zz}\right) c_z (s_x+i s_y)&\dots\\
\dots&4 \eta^{v2}_{xz} (c_y s_x-i c_x s_y)-\left(\sqrt{3} \eta^{v2}_{xx}+\eta^{v2}_{xz}-3 \eta^{v2}_{zx}-\sqrt{3} \eta^{v2}_{zz}\right) c_z (s_x-i s_y) &   \left(\sqrt{3} \eta^{v2}_{xx}-\eta^{v2}_{xz}+3 \eta^{v2}_{zx}-\sqrt{3} \eta^{v2}_{zz}\right) (c_x-c_y) s_z &\dots\\
\dots&\left(3 \eta^{v2}_{xx}-\sqrt{3}   \eta^{v2}_{xz}-\sqrt{3} \eta^{v2}_{zx}+\eta^{v2}_{zz}\right) (c_x+c_y) s_z & 4 \eta^{v2}_{zz} (c_y s_x-i c_x s_y)+\left(3 \eta^{v2}_{xx}+\sqrt{3}   \eta^{v2}_{xz}+\sqrt{3} \eta^{v2}_{zx}+\eta^{v2}_{zz}\right) c_z (s_x-i s_y) &\dots\\
\dots&4 \eta^{v2}_{zz} (c_y s_x+i c_x s_y)+\left(3 \eta^{v2}_{xx}+\sqrt{3} \eta^{v2}_{xz}+\sqrt{3} \eta^{v2}_{zx}+\eta^{v2}_{zz}\right) c_z (s_x+i s_y)&   -\left(3 \eta^{v2}_{xx}-\sqrt{3} \eta^{v2}_{xz}-\sqrt{3} \eta^{v2}_{zx}+\eta^{v2}_{zz}\right) (c_x+c_y) s_z&\dots
\end{smallmatrix}
\eea
\bea
\left.
\begin{smallmatrix}
\dots&-2 \left(\eta^{v2}_{x7}+\sqrt{3} \eta^{v2}_{z7}\right) (c_x+c_y) s_z & 4 \eta^{v2}_{x7} (c_y s_x-i c_x s_y)-2 \left(\eta^{v2}_{x7}-\sqrt{3} \eta^{v2}_{z7}\right) c_z (s_x-i s_y) \\
\dots&4 \eta^{v2}_{x7} (c_y s_x+i c_x s_y)-2   \left(\eta^{v2}_{x7}-\sqrt{3} \eta^{v2}_{z7}\right) c_z (s_x+i s_y) & 2 \left(\eta^{v2}_{x7}+\sqrt{3} \eta^{v2}_{z7}\right) (c_x+c_y) s_z \\
\dots&2 \left(\sqrt{3}   \eta^{v2}_{x7}-\eta^{v2}_{z7}\right) (c_x-c_y) s_z & 4 \eta^{v2}_{z7} (c_y s_x+i c_x s_y)+2\left(\sqrt{3} \eta^{v2}_{x7}+\eta^{v2}_{z7}\right) c_z   (s_x+i s_y) \\
\dots&4 \eta^{v2}_{z7} (c_y s_x - i c_x s_y)+\left(\sqrt{3} \eta^{v2}_{x7}+\eta^{v2}_{z7}\right) c_z (s_x-i s_y) & -2 \left(\sqrt{3}   \eta^{v2}_{x7}-\eta^{v2}_{z7}\right) (c_x-c_y) s_z
\end{smallmatrix}
\right)
\eea
with $\eta^{v2}_{xx}=-0.232$, $\eta^{v2}_{xz}=-0.152$, $\eta^{v2}_{zx}=0.106$, $\eta^{v2}_{zz}=0.589$, $\eta^{v2}_{x7}=-0.143$, $\eta^{v2}_{z7}=0.506$.

%


\subsection{Minimal model}

Considering the complexity of the model described so far, we note that not all the reported parameters are required for a qualitative description of {\sm} and {\pu}.
A minimal model entailing all the orbitals and correctly reproducing the bandstructure must contain:
(i) the onsite energies $\epsilon^d$, $\epsilon_{\Gamma_8}^f$,
$\epsilon_{\Gamma_7}^f$,
(ii) the first NN hoppings $\eta_z^{d1}$, $\eta_z^{f1}$,
(iii) the second NN hoppings $\eta_z^{d2}$, $\eta_z^{f2}$, $\eta_7^{f2}$,
(iv) the third NN hopping $\eta^{f3}_7$ -- this connects the $\Gamma_7$ orbitals via a term $(-8\tf\eta^{f3}_7c_xc_yc_z)$ with $\eta^{f3}_7=1.25$ -- and
(v) the second NN hybridizations $\eta^{v2}_{zz}$, $\eta^{v2}_{z7}$.

With respect to Ref. \onlinecite{takimoto}, where parameters $t_d$,
$t_d'$, $t_f$, $t_f'$ were used, we have:
$t_d \propto \bar{t}_d \eta_z^{d1}$, $t'_d \propto \bar{t}_d
\eta_z^{d2}$, $t_f \propto \bar{t}_f \eta_z^{f1}$, $t'_f \propto
\bar{t}_d \eta_z^{f2}$.
The main difference with the aforementioned paper, apart form the
inclusion of the $\Gamma_7$ orbital, is the second NN hybridization,
which is unusually more important than the first NN one.


\section{Slave-boson mean-field approximation}

To account for the strong interaction $H_U$ of the Anderson lattice model, Eq.~{\eqh} of the main text, we utilize the popular slave-boson approach.\cite{read_newns_slavebosons_1,read_newns_slavebosons, hewson}
In its standard formulation, it is designed to implement a Hubbard-like repulsion of strength $U$ in the limit $U\rightarrow \infty$, where all states with more than one electron on each $f$ orbital are forbidden. The remaining states of the local $f$ Hilbert space are represented by auxiliary particles, with $b_i$ for empty ($f^0$) and $\tilde{f}_{i\alpha\sigma}$ for singly occupied ($f^1$) orbitals on site $i$, such that ${f}_{i\alpha\sigma}= b_i^\dagger \tilde{f}_{i\alpha\sigma}$.
The Hilbert space is constrained by $b_i^\dagger b_i +\sum_{\alpha\sigma} \tilde{f}^\dagger_{i\alpha\sigma}\tilde{f}_{i\alpha\sigma}=1$.
It is convenient to choose $b_i$ bosonic and $\tilde{f}_{i\alpha\sigma}$ fermionic, and to employ a saddle-point approximation $b_i \rightarrow b=\langle b_i \rangle$.
With fluctuations of $b_i$ frozen, the above constraint is imposed in a mean-field fashion using a Lagrange multiplier $\lambda$. This eventually reduces the Anderson model to a model of non-interacting, but interaction-renormalized, bands.
Together with the global chemical potential $\mu$, there are three  parameters $b$, $\lambda$, $\mu$ which need to be determined self-consistently.

In the present case, the electronic configuration of {\sm} and {\pu} is mixed valent, $d^1 f^5  \leftrightarrow d^0 f^6$. Hence, the infinite repulsion suppresses states with less than five $f$ electrons per site, and it is convenient to work in a {\em hole} representation.
Formally, we perform a particle--hole transformation on both $f$ and $d$ orbitals, such that $d^0 f^6$ becomes $d^4 f^0$ (no $f$ holes), and $d^1 f^5$ becomes $d^3 f^1$, i.e., the $f$ state with single (hole) occupancy. Then, the slave-boson method can be applied as before.


In situations with full translation symmetry the resulting mean-field Hamiltonian can be written in momentum space and takes the form
\be\label{h_mf}
H_\bk^{\rm MF}=H_{dd}^{\rm MF}+H_{df}^{\rm MF}+H_{ff}^{\rm MF},
\ee
with its pieces in hole representation
\bea
H_{dd}^{\rm MF}&=&\sum_{\bk\sigma\alpha\alpha'} [(-\epsilon^d_\alpha+\mu) d^\dagger_{\bk\sigma\alpha}d_{\bk\sigma\alpha}+t^d_{\bk\sigma\alpha\alpha'} (d^\dagger_{\bk\sigma\alpha}d_{\bk\sigma\alpha'}+h.c.)],\\
H_{ff}^{\rm MF}&=&\sum_{\bk\sigma\sigma'\alpha\alpha'} [(-\epsilon^f_{\alpha}+\mu+\lambda)\tilde f^\dagger_{\bk\sigma\alpha}\tilde f_{\bk\sigma\alpha}-b^2(t^f_{\bk\sigma\sigma'\alpha\alpha'} \tilde f^\dagger_{\bk\sigma\alpha}\tilde f_{\bk\sigma'\alpha'}+h.c.)],\\
H_{df}^{\rm MF}&=&-b\sum_{\bk\sigma\sigma'\alpha\alpha'} (V_{\bk\sigma\sigma'\alpha\alpha'} 	d^\dagger_{\bk\sigma\alpha}\tilde f_{\bk\sigma'\alpha'}+h.c.).
\eea
Here $\bk\equiv(k_x,k_y,k_z)$ is a momentum in the first Brillouin zone (BZ) $-\pi\leq k_x,k_y,k_z<\pi$,
$t^d_{\bk\sigma\alpha\alpha'}$, $t^f_{\bk\sigma\sigma'\alpha\alpha'}$ and $V_{\bk\sigma\sigma'\alpha\alpha'}$ are the Fourier transforms of the hopping parameters.
The self-consistent equations to determine $\mu$, $b$, and $\lambda$ read
\begin{eqnarray}
\label{mf1}
N_e &=&\sum_{\bk\sigma\alpha} (\langle d_{\bk\sigma\alpha}^\dagger d_{\bk\sigma\alpha}\rangle + \langle \tilde{f}_{\bk\sigma\alpha}^\dagger \tilde{f}_{\bk\sigma\alpha}\rangle),\\
\label{mf2}
0 &=& 2b\left(-\frac{1}{N_s}\sum_{\bk\sigma\alpha} t^f_{\bk\sigma\sigma'\alpha\alpha'} \langle \tilde{f}_{\bk\sigma\alpha}^\dagger \tilde{f}_{\bk\sigma'\alpha'}\rangle+\lambda \right)-\frac{1}{N_s}\sum_{\bk\sigma\alpha\sigma'\alpha'}\left( V_{\bk\sigma\alpha\sigma'\alpha'}\langle d_{\bk\sigma\alpha}^\dagger \tilde{f}_{\bk\sigma'\alpha'}\rangle+h.c.\right),\\
\label{mf3}
1 &=& b^2+\frac{1}{N_s}\sum_{\bk\sigma\alpha}{\langle \tilde{f}_{\bk\sigma\alpha}^\dagger \tilde{f}_{\bk\sigma\alpha}\rangle},
\end{eqnarray}
where $N_s$ is the number of lattice sites. The filling corresponding to the Kondo insulator is given by $N_e=4 N_s$, meaning 4 holes, i.e., 6 electrons, per site.
More details can be found in Ref. \onlinecite{prb_io_tki}.

To determine the mean-field parameters for our model, we have solved the equations \eqref{mf1}, \eqref{mf2}, and \eqref{mf3} iteratively at a temperature $T$ of $10^{-4}$~eV using a momentum-space grid with $25^3$ points.
We obtain $\lambda=0.58$~eV, $\mu=0.10$~eV, $b=0.72$, the latter value implying a mixed-valence situation with $n_f\approx5.5$. This value agrees well with the experimentally determined $f$ valence of {\sm}.\cite{mizumaki}

\end{widetext}


\section{Scattering matrix}

To calculate impurity-induced changes of electron propagators, we start from the Green's function $\hat{G^0}$ of a clean slab with a periodic boundary conditions along $x$ and $y$ and open boundary conditions along $z$. $\hat{G^0}$ is diagonal in the in-plane momentum $\bk=(k_x,k_y)$ and can be calculated according to
\begin{equation}\label{gkslab}
\hat{G^0}_{za,z'a'}(E,\bk)=\left(\hat{1}(E+\mu+i\delta)-\hat{H}^{\rm MF}_{\bk}\right)^{-1}_{za,z'a'},
\end{equation}
with $\hat{H}^{\rm MF}_{\bk}$ being the mean-field Hamiltonian from Eq.~\eqref{h_mf} after Fourier transformation w.r.t. the in-plane coordinates. Here, $a$ and $a'$ are orbital indices with $1\le a,a'\le 10$, $\mu$ is the chemical potential, and $\delta$ is an artificial broadening parameter.

The effect of an isolated impurity is obtained using the standard T-matrix formalism,
\begin{equation}
\hat{G}(E)=\hat{G^0}(E)+\hat{G^0}(E)\hat{T}(E)\hat{G^0}(E),\label{gg0t}
\end{equation}
where the scattering T matrix is determined as
\begin{equation}
\hat{T}(E)=\hat{V}\left(1-\hat{G^0}(E)\hat{V}\right)^{-1}.
\end{equation}
Here, all matrices depend on the real-space positions $\br=(x,y,z)$ and $\br'=(x',y',z')$, with $1\le x,y,x',y' \le N_x$, $1\le z,z' \le N_z$, and on orbital indices $a$ and $a'$; the real-space form of $\hat{G^0}$ is obtained from Eq.~\eqref{gkslab} by fast Fourier transformation.
The scattering potential $V$ is non-zero on the impurity site only: for Kondo holes, we take an on-site $f$ energy of $V=100$~eV, while for weak scatterers we modify the on-site energy in one of the orbitals by $V=10$~meV.
Our slab thickness is $N_z=25$. To reach sufficient energy resolution, we have used $N_x=801$, $\delta=1$~meV.

For further technical details, we refer the reader to Ref.~\onlinecite{prb_io_tki} where the same approach was used to study a simpler four-orbital model for tetragonal topological Kondo insulators.


\section{STS signal}

Here we summarize the calculation of the STS signal which involves a modelling of the electronic tunneling processes between the microscope tip and the material's $(001)$ surface. As will become clear below, an important ingredient is the orbital character of the electronic states in the tip.

\begin{figure}[tbp]
\includegraphics[width=0.18\textwidth]{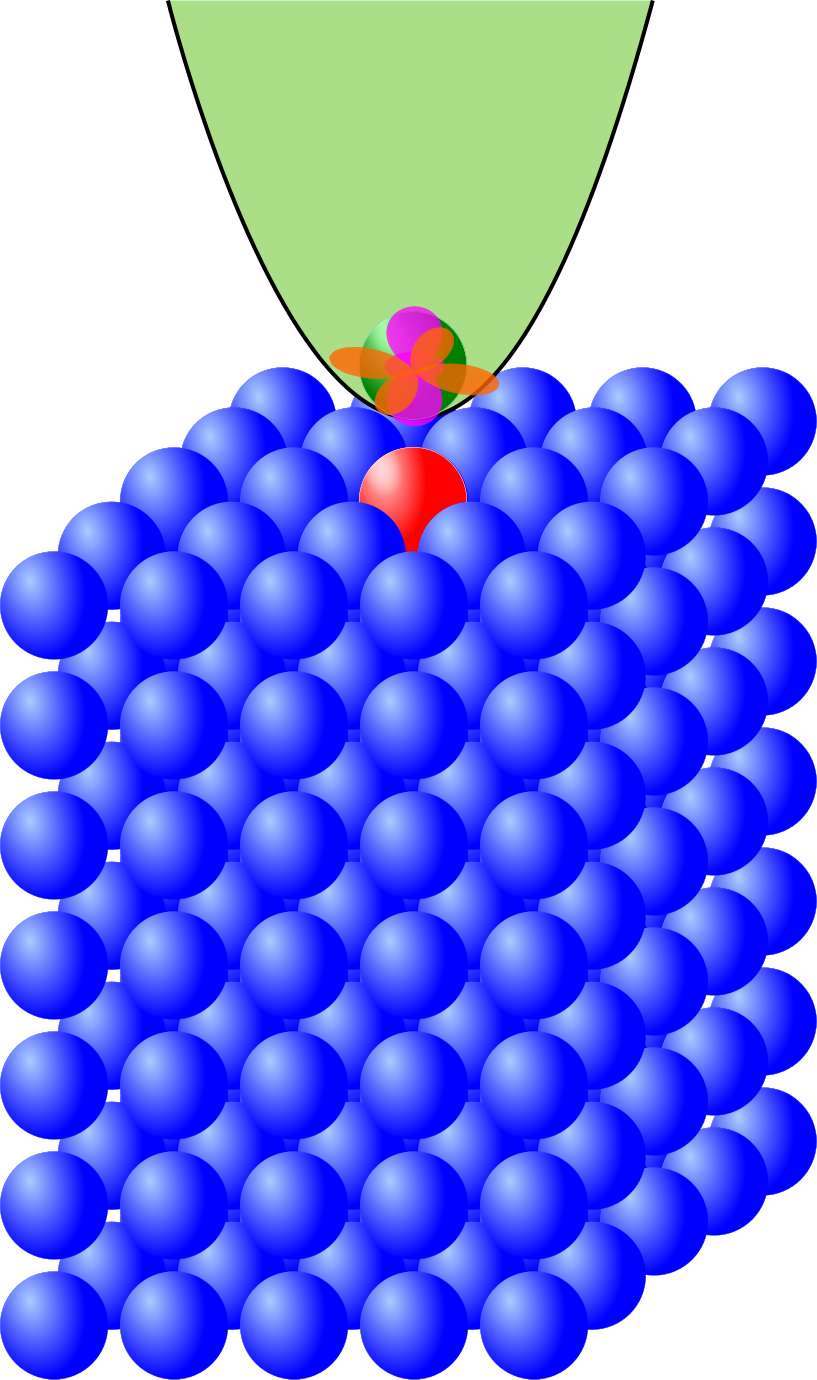}
\caption{Schematic picture of our setup: in the surface layer of a slab of Sm atoms (blue), an impurity or Kondo hole (red) is introduced.
Above the slab we put an STS tip (green) ending with a single atom, which has two conduction channels (both of them spin degenerate), one $d_{x^2-y^2}$-like (orange), and one $d_{z^2}$-like (purple).
}
\label{fig_stm_slab}
\end{figure}

We assume a tip ending with a single apex atom and a vertical tunneling path between this tip atom and a Sm (Pu) atom beneath it, Fig.~\ref{fig_stm_slab}. Non-zero tunneling matrix elements arise only for tip states whose wavefunction symmetry, projected into the $xy$ plane, is $s$-like or $d_{x^2-y^2}$-like. Taking the spin degree of freedom into account, the modelling thus requires four tip-electron channels.
For the purpose of numerical estimates (see below), we assume the tip states to be those of $d$ electrons, and hence consider tip electrons in $d_{x^2-y^2}$ and $d_{z^2}$ orbitals (the latter has an $s$-like wavefunction when projected into the $xy$ plane).
By extending the treatment of Ref. \onlinecite{dzero_cotunneling} to multichannel transport, and restricting ourselves to vertical tunneling, the tunneling Hamiltonian can be written as
\be
H_T=\sum_{\substack{\sigma=\uparrow, \downarrow \\ \alpha=d_{x^2-y^2},d_{z^2}}} (p^\dagger_{\sigma\alpha} \psi_{\sigma\alpha}  +h.c.),
\ee
where  we assume operators $p_{\sigma\alpha}$ to describe tip orbitals and
\be
\psi_{\sigma\alpha}=\sum_{\sigma'\alpha'}( \bar{t}^d_{\sigma\alpha\sigma'\alpha'} d_{\sigma'\alpha'}+b \bar{t}^f_{\sigma\alpha\sigma'\alpha'} \tilde{f}_{\sigma'\alpha'})
\ee
is built with hopping matrix elements from the tip to the surface, together with surface orbitals of the atom beneath the tip. Note that the physical $f$-electron operator has been expressed as $b\tilde{f}$, with the slave-boson renormalization factor $b$ entering.\cite{balatsky_kondohole}

The differential conductance $g(E)$ is now the sum of four terms, each proportional to the imaginary part of $G_{\psi_{\sigma\alpha}}$ (the ``cotunneling'' Green's function for operator $\psi_{\sigma\alpha}$), times the corresponding density of states of the tip $\rho^{TIP}_{\sigma\alpha}$, that for simplicity we take energy-, spin-, and orbital-independent:
\be
g(E)=-\frac{2e^2}{\hbar} \rho^{TIP} \!\! \sum_{\substack{\sigma=\uparrow, \downarrow \\ \alpha=d_{x^2-y^2},d_{z^2}}} \I G_{\psi_{\sigma\alpha}}(E).
\ee

We now write $\psi_{\sigma\alpha}$ as a $4\times 10$ $\psi$ matrix, on the basis of the four tip orbitals and the 10 surface orbitals per site, whose (real) coefficients, constrained by symmetry, are
\be
\psi=\left( \begin{array}{cccccccccc}
\bar{t}^d_1 & 0 & 0 & 0 & b\bar{t}^f_1 & 0 & 0 & 0 & b\bar{t}^f_3 & 0\\
0 & \bar{t}^d_1 & 0 & 0 & 0 & -b\bar{t}^f_1 & 0 & 0 & 0 & -b\bar{t}^f_3\\
0 & 0 & \bar{t}^d_2 & 0 & 0 & 0 & b\bar{t}^f_2 & 0 & 0 & 0\\
0 & 0 & 0 & \bar{t}^d_2 & 0 & 0 & 0 & -b\bar{t}^f_2 & 0 & 0
\end{array} \right)
\ee
where
$\bar{t}^d_1\equiv \bar{t}^d_{x^2-y^2\sigma, x^2-y^2\sigma}$,
$\bar{t}^d_2\equiv \bar{t}^d_{z^2\sigma, z^2\sigma}$,
$\bar{t}^f_1\equiv \bar{t}^f_{x^2-y^2\uparrow, \Gamma_8^{(1)}+}$,
$\bar{t}^f_2\equiv \bar{t}^f_{z^2 \uparrow, \Gamma_8^{(2)}+}$, and
$\bar{t}^f_3\equiv \bar{t}^f_{x^2-y^2\uparrow, \Gamma_7+}$
represent the effective coupling of each orbital to the tip.
To compute the tunneling conductance, we need the local Green's function of the material,  $G_{\sigma\alpha\sigma'\alpha'}(E, \br, \br)$, which formally is a $10\times 10$ matrix whose entries are computed numerically through the scattering-matrix approach, Eq.~\eqref{gg0t}.
We finally find:
\be\label{tmGtm}
g(E)=-\frac{2e^2}{\hbar} \rho^{TIP}\I\Tr [\hat\psi \hat G(E)  \hat\psi^T].
\ee

The parameters $\bar{t}^d_1$, $\bar{t}^d_2$, $\bar{t}^f_1$, $\bar{t}^f_2$, $\bar{t}^f_3$ in our theory are free, and will depend on the details of the tip-surface coupling.
To fix them in an approximate way using available data, we make the following assumption: the tip apex atom has the {\em same} $d$ shell as Sm or Pu. If this tip atom has a distance to the topmost surface atom identical to the material's lattice spacing, the hopping parameters can be copied from our original tight-binding Hamiltonian:
$\bar{t}^d_1=t^d_{\sigma x^2-y^2 x^2-y^2}=0.09$~eV, $
\bar{t}^d_2=t^d_{\sigma z^2 z^2}=-0.81$~eV,
$\bar{t}^f_1=V_{ x^2-y^2\uparrow \Gamma_8^{(1)}+}=-0.04$~eV,
$\bar{t}^f_2=V_{ z^2\uparrow \Gamma_8^{(2)}+}=0.21$~eV,
$\bar{t}^f_3=V_{ x^2-y^2\uparrow \Gamma_7+}=0.02$~eV.
These parameters have been used for our figures which show $\rho_{STS}(E)=-1/\pi \I \sum_{\sigma\alpha} G_{\psi_{\sigma\alpha}}(E)$.
We note that the leading effect of varying the distance between tip and surface is a simple rescaling of all hopping matrix elements, such that the total signal needs to be multiplied by a distance-dependent constant.

With this choice of tunneling parameters, most of the signal comes from $d_{z^2}$ and $\Gamma_8^{(2)}$ orbitals of the material, which are the ones which extend mostly in the $z$ direction and have an $s$-like projection such that they effectively couple to the $d_{z^2}$ orbital of the tip.
If we would ignore the contributions from the remaining orbitals, $d_{x^2-y^2}$, $\Gamma_8^{(1)}$, and $\Gamma_7$, we would recover the simple conduction model with one $d$ (or $s$) and one $f$ orbital of Ref.~\onlinecite{dzero_cotunneling}, with $\tilde{t}_f/t_c\equiv b\bar{t}_{2}^f/\bar{t}_{2}^d=-0.19$.
It is worth emphasizing that even orbitals which do not directly couple to the tip are nevertheless important for the physics of the microscopic model; this is shown explicitly in the following sections.

\begin{widetext}

\section{Expectation value of the spin}
\label{sec:spin}

In this section we describe the calculation to determine the spin structure of the topological surface states which can be measured using spin-resolved ARPES experiments such as the one in Ref.~\onlinecite{smb6_arpes_mesot_spin}.

We first start with the spin-integrated ARPES signal. Given the Green's function $\hat{G}^0_{za,z'a'}(E,\bk)$ from Eq. \eqref{gkslab},
the surface ARPES signal of Fig. 4(a,b,c) of the main text and of Fig. \ref{fig_qpimod}(a),(c),(e) below is obtained through
\begin{align}
A(E, \bk, z=1)&=-\frac{1}{\pi}\I \sum_{az=1}  G^0_{za,za}(E,\bk) 
=-\frac{1}{\pi}\I\Tr[\hat G^0(E,\bk) \hat Z],
\end{align}
where the operator $\hat Z$ is a projector on the subspace with $z=1$:
\be
\hat Z_{za,z'a'}=\delta_{z=z'=1}.
\end{equation}

In analogy, the intensity of the spin-polarized ARPES signal at energy $E$ and in-plane momentum $\bk$ obtained from layer $z=1$ is
\be
\label{spinex}
\langle \vec{\sigma} \rangle(E,\bk)=-\frac{1}{\pi}\I \Tr [\hat G^0(E,\bk) \vec\sigma \hat Z],
\ee
this quantity corresponds to the spin expectation of the ejected electron.
We observe that ($a\equiv{\beta\alpha\sigma}$)
\be
\langle z \beta\alpha\sigma|\vec\sigma \hat Z|z' \beta'\alpha' \sigma'\rangle=\delta_{z=z'=1}\delta_{\beta\beta'}\langle \sigma \alpha|\vec\sigma |\sigma' \alpha'\rangle,
\end{equation}
so we only need to compute matrix elements $\langle \sigma \alpha|\vec\sigma |\sigma' \alpha'\rangle$, where,
if $\beta=c$, $\alpha,\alpha'=d_{x^2-y^2}/d_{z^2}$, and $\sigma,\sigma'=\uparrow/\downarrow$ ,
while,
if $\beta=f$, $\alpha,\alpha'=\Gamma_8^{(1)}/\Gamma_8^{(2)}/\Gamma_7$, and $\sigma,\sigma'=+/-$ .
The non-zero matrix elements for $d$ states are trivially:
\begin{eqnarray}
\langle d \alpha \uparrow | \sigma^x|d \alpha' \downarrow\rangle=&\langle d \alpha \downarrow | \sigma^x|d \alpha' \uparrow\rangle=&\delta_{\alpha\alpha'},\\
\langle d \alpha \uparrow | \sigma^y|d \alpha' \downarrow\rangle=&\langle d \alpha \downarrow | \sigma^y|d \alpha' \uparrow\rangle^*=&-i\delta_{\alpha\alpha'} ,\\
\langle d \alpha \uparrow | \sigma^z|d \alpha' \uparrow\rangle=&-\langle d \alpha \downarrow | \sigma^z|d \alpha' \downarrow\rangle=&\delta_{\alpha\alpha'}.
\end{eqnarray}
To obtain the expectation value of the spin on $f$ states
we trace out the orbital degree of freedom; in the basis $\Gamma_8^{(1)}+,\Gamma_8^{(1)}-,\Gamma_8^{(2)}+,\Gamma_8^{(2)}-,\Gamma_7+,\Gamma_7-$ we get :
\begin{align}
&\langle f \alpha\sigma  | (\sigma^x,\sigma^y, \sigma^z)|f \alpha'\sigma' \rangle=&\nonumber\\
&\frac{1}{21}\left( \begin{array}{llll|ll}\label{table_spin}
11(0,0,-1) & 5(-1,i,0) & (0,0,0) & 2\sqrt{3}(-1,-i,0) & {4\sqrt{5}}(0,0-1)&{2\sqrt{5}} (1,-i,0)\\
5(-1,-i,0) & 11 (0,0,1) & 2\sqrt{3} (-1,i,0) & (0,0,0)&2\sqrt{5}(1,i,0)&{4\sqrt{5}}(0,0,1)\\
(0,0,0) & 2\sqrt{3}(-1,-i,0) & 3 (0,0,-1) & 9(-1,i,0)& (0,0,0)& {2}\sqrt{15}(1,i,0)\\
{2}\sqrt{3}(-1,i,0)&(0,0,0)&9 (-1,-i,0)&3 (0,0,1) &{2}\sqrt{15}(1,-i,0)&(0,0,0)\\\hline
{4\sqrt{5}}(0,0,-1)&{2\sqrt{5}} (1,-i,0) &(0,0,0)&{2}\sqrt{15}(1,i,0)&{5}(0,0,1)&{5}(1,-i,0)\\
{2\sqrt{5}}(1,i,0)&{4\sqrt{5}}(0,0,1)&{2}\sqrt{15}(1,-i,0)&(0,0,0)&{5}(1,i,0)& {5}(0,0,-1)
  \end{array}\right)&.
\end{align}

\begin{figure}[tbp]
\includegraphics[width=0.95\textwidth]{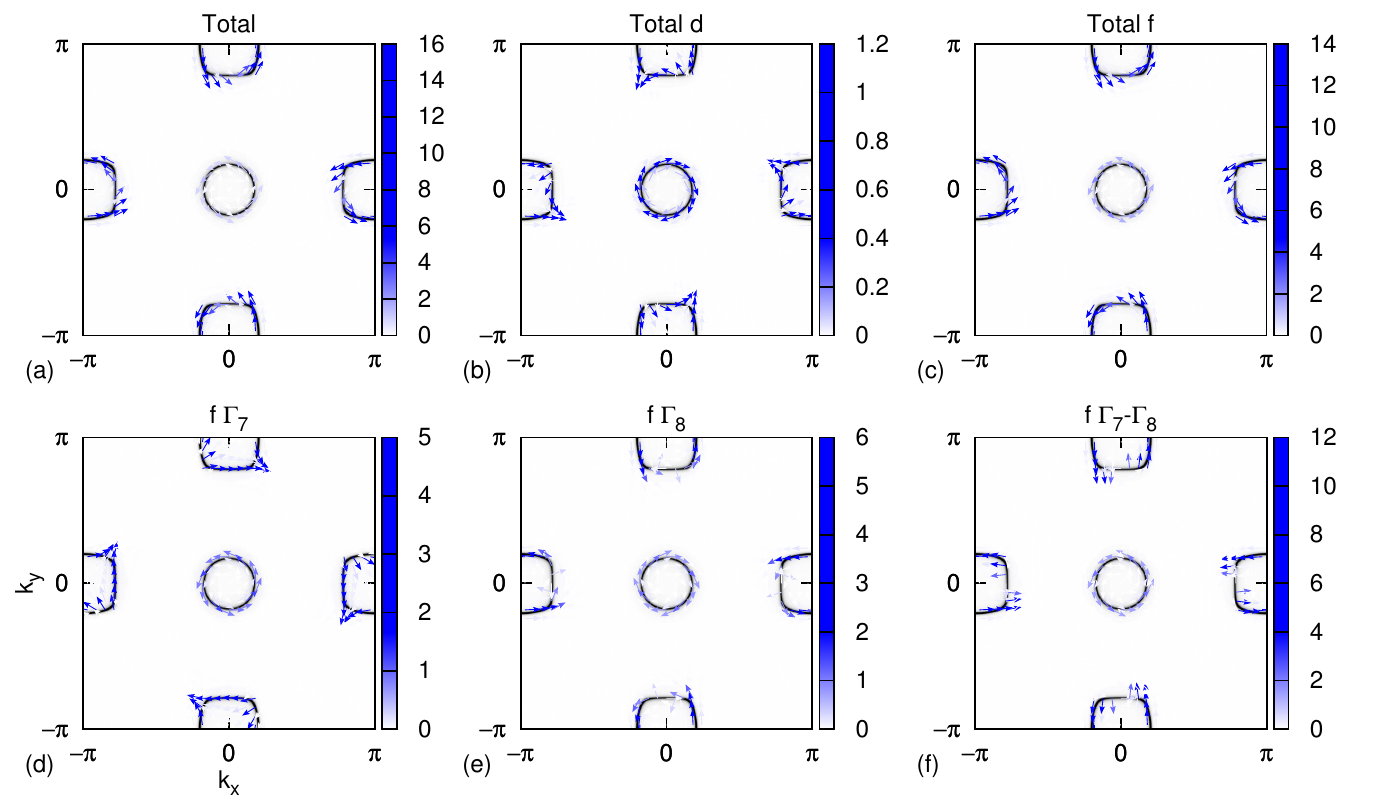}
\caption{Expectation value of the spin, $\langle \vec{\sigma} \rangle(E,\bk)$ \eqref{spinex}, on the first layer at the Fermi energy, computed with $N_x=801$, $\delta=1$~meV and averaged over a 33$\times$33 grid. The different panels show
(a) the total expectation value,
(b) the contribution from $d$ states,
(c) the contribution from $f$ states,
(d) the contribution from $f$ $\Gamma_7$ states [lower diagonal block of Eq.~\eqref{table_spin}],
(e) the contribution from $f$ $\Gamma_8$ states [upper diagonal block of Eq.~\eqref{table_spin}],
(f) the mixed $\Gamma_7$--$f$ $\Gamma_8$ contribution [off-diagonal blocks of Eq.~\eqref{table_spin}].
The arrows show the in-plane spin direction (the out-of-plane component is negligible); the color code indicates the magnitude of the signal.
}
\label{fig_spin}
\end{figure}

\end{widetext}

In Fig.~\ref{fig_spin} we report the results of this calculation. Panel (a) shows $\langle \vec{\sigma} \rangle(E,\bk)$ at fixed $E=0$ as function of $\bk$ -- for sharp quasiparticles this signal is only non-zero at the iso-energy contours (for numerical reasons we have used a finite broadening). Importantly, the result in Fig.~\ref{fig_spin}(a) is consistent with the corresponding experimental result obtained on {\sm}.\cite{smb6_arpes_mesot_spin}

The remaining panels of Fig.~\ref{fig_spin} illustrate the different orbital contributions to $\langle \vec{\sigma} \rangle(E\!=\!0,\bk)$, obtained by only taking into account a partial set of spin matrix elements.
It is remarkable that the different orbitals yield {\em qualitatively} distinct contributions to $\langle \vec{\sigma} \rangle(E,\bk)$: The winding of the in-plane spin components along an iso-energy contour is opposite in panels (d) and (e) for both the $\bar\Gamma$ and the $\bar X$ cones. This underlines that the observable spin structure depends sensitively on the orbital content of the surface states, as mentioned in the main text.


\section{Additional QPI results}

This section contains additional results for quasiparticle interference (QPI) spectra for the microscopic model discussed in the paper and its variants.

\subsection{Full model: Momentum-space cuts}

\begin{figure}[tbp]
\includegraphics[width=0.49\textwidth]{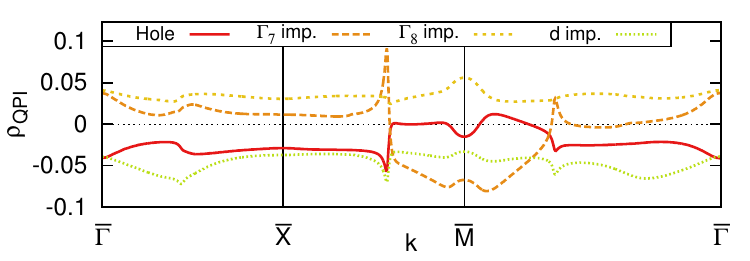}
\includegraphics[width=0.49\textwidth]{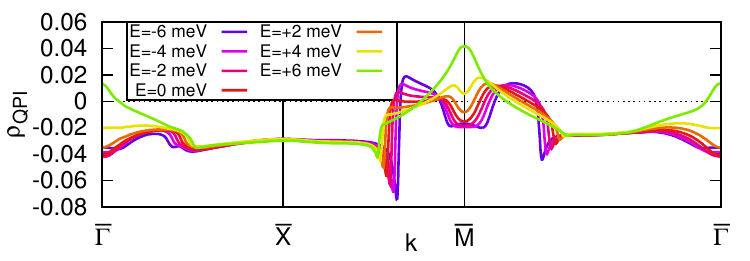}
\caption{(a) QPI signal at the Fermi energy, $\rho_{QPI}(E=0)$, along the $\bar\Gamma \bar{X} \bar{M} \bar\Gamma$ path in the 2D Brillouin zone, comparing a Kondo hole, a weak $\Gamma_7$ and a weak $\Gamma_8$ impurity ($V=10$~meV, signal is multiplied by 20), and a weak $d$ impurity ($V=10$~meV, signal is multiplied by 200).
(b) QPI signal for a Kondo hole at different energies.
}
\label{fig_qpi_1d}
\end{figure}

Fig.~\ref{fig_qpi_1d} displays QPI data as in Fig.~4 of the main paper, but here $\rho_{QPI}$ is shown along a path in the surface Brillouin zone. Panel \ref{fig_qpi_1d}(a) shows the same data as in Figs.~4(e), (h), and (k), plus the signal for a weak scatterer placed in the $d$ band. Panel \ref{fig_qpi_1d}(b) focusses in the Kondo-hole case and displays the energy evolution of the corresponding QPI signal.

All curves are essentially flat near $\bar\Gamma$, corresponding to suppressed intracone scattering. As mentioned in the main text, the four cases, however, differ in the behavior near $\bar M$: whereas the $\Gamma_8$ scatterer produces no appreciable signal from intercone scattering, the other cases lead to intercone scattering peaks which are strong both for the Kondo hole and for the $\Gamma_7$ scatterer. The origin is in the intricate spin structure of the Dirac-cone states, as discussed in Sec.~\ref{sec:spin} above.

\subsection{Comparison of full and reduced models}

\begin{figure}[!t]
\includegraphics[width=0.44\textwidth]{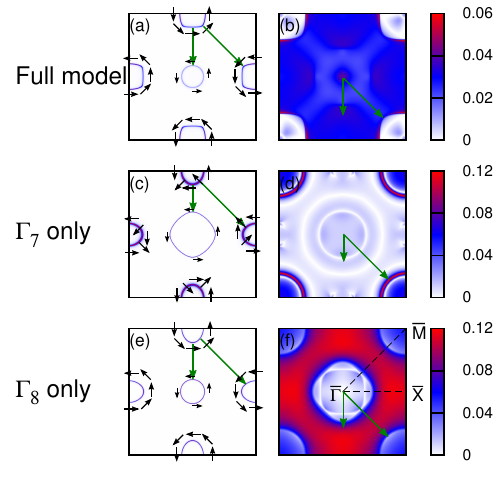}
\caption{Surface ARPES and QPI signals inside the bulk gap for
(a,b) the full model at $E=0$,
(c,d) the ``$\Gamma_7$ only'' model at $E=-14$~meV, such that $E_{\bar\Gamma}<E<E_{\bar X}$, and
(e,f) the ``$\Gamma_8$ only'' at $E=+50$~meV where $E>E_{\bar\Gamma},E_{\bar X}$.
In each ARPES figure we schematically show the expectation value of the spin.
The QPI signal is shown as $|\rho_{QPI}|$ and has been calculated for scattering off isolated Kondo holes. For details see text.
}
\label{fig_qpimod}
\end{figure}

\begin{figure}[b]
\includegraphics[width=0.49\textwidth]{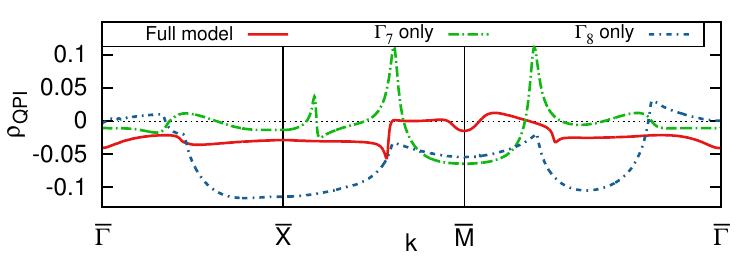}
\caption{QPI signal along the $\bar\Gamma \bar{X} \bar{M} \bar\Gamma$ path in the 2D Brillouin zone for a Kondo hole, comparing the full model, the ``$\Gamma_7$ only'' model, and the ``$\Gamma_8$ only'' model. The curves are shown for the same energies as in Fig.~\ref{fig_qpimod}.
}
\label{fig_qpimod_1d}
\end{figure}

As announced in the main text, we have also considered orbitally reduced versions of the model, obtained by retaining only the $\Gamma_7$ doublet or the $\Gamma_8$ quartet in the model Hamiltonian {\eqh} of the main text. Both cases give rise to a TKI with three Dirac cones at $\bar\Gamma$ and $\bar{X}$, qualitatively similar to the full model. However, the magnitude of the bulk gap changes significantly (see also Fig.~5 of Ref.~\onlinecite{pub6}): For the ``$\Gamma_7$ only'' model we find the bulk gap between $- 25$~meV and $25$~meV, while for the ``$\Gamma_8$ only'' model the gap range is $[- 80,80]$~meV.
Using these reduced models, we have determined the ARPES and QPI signals as well as the spin structure of the Dirac cones, with results and their comparison to that of the full model shown in Figs.~\ref{fig_qpimod} and \ref{fig_qpimod_1d}.

In the ``$\Gamma_8$ only'' model the spin structure is similar to the one of the full model, Figs.~\ref{fig_qpimod}(a) and (e). Consequently, the QPI signal from intercone scattering is similarly flat, i.e., non-peaked, for $\bar \Gamma$--$\bar X$ scattering and only weakly peaked for $\bar X$--$\bar X'$ scattering -- this is particularly clear in Fig.~\ref{fig_qpimod_1d}. Note, however, that the detailed momentum-space distribution of QPI intensity is nevertheless rather different in the two cases.

In contrast, the ``$\Gamma_7$ only'' case offers an opposite scenario: The winding of the in-plane spin component of the Dirac cones at $\bar X$ and $\bar X'$ is reversed, such that the expectation value of the spin is now roughly parallel for pairs of stationary points, Figs.~\ref{fig_qpimod}(c). This result closely resembles the one of Ref.~\onlinecite{yu_kp_smb6}, where a LDA+Gutzwiller approach was used to compute the surface states of {\sm} and their spin structure. (We recall that the experimental ARPES results\cite{smb6_arpes_mesot_spin} are different and instead agree with our full calculation.) This distinctly different spin structure in turn leads to a sharp QPI peak corresponding to intercone $\bar X$--$\bar X'$ scattering, Fig.~\ref{fig_qpimod_1d}.
These findings also explain how the relative $\Gamma_7$/$\Gamma_8$ weight on the surface states, and in particular on the $\bar X$ cones, controls the strength of the $\bar X$--$\bar X'$ scattering peak,
in addition to the relative $\Gamma_7$/$\Gamma_8$ weight of the impurity, as shown in the main text.

We note that we have chosen, for illustration purposes, an energy with $E_{\bar\Gamma}<E<E_{\bar X}$ in Figs.~\ref{fig_qpimod}(c,d). (For the other models, this energy interval has no overlap with the bulk gap and hence cannot be probed by surface-state QPI.) As a result, the spin on one of the two cones is reversed, and hence a peak is expected for scattering between the $\bar \Gamma$ and $\bar X$ cones. Such a peak, albeit weak, is indeed seen in Figs.~\ref{fig_qpimod}(d) and \ref{fig_qpimod_1d}.
In all other cases, $\bar\Gamma$ -- $\bar X$ scattering does not induce a sizeable QPI signal.

\subsection{QPI summary}

Let us quickly summarize our insights concerning the QPI signal arising from topological surface states with multiple inequivalent Dirac cones. Most generally, we find that:
(i) intracone scattering generically gives rise to weak and flat (non-peaked) contributions, and
(ii) intercone scattering can lead to either strong and distinctly peaked signals or to weak flat signals, depending on the (relative) spin structure of the cones.

For identical cones, such as the $\bar{X}$ ones in {\sm}, two limiting scenarios concerning intercone scattering are possible: one in which the spin for pairs of stationary points is parallel which leads to a QPI peak, and one in which this spin is antiparallel which leads to a QPI plateau.
However, the multi-orbital nature of the underlying model allows for departures from these limiting cases: The contributions to spin (or other quantum numbers distinguishing Kramers-degenerate partners of states) from the different orbitals can be qualitatively different, see Fig.~\ref{fig_spin} above, such the orbital content of both surface states and scatterers eventually determine the structure of the QPI signal, and QPI peaks may occur even if the spin structure (as detected by spin-resolved ARPES) would suggest otherwise.

For scattering between nonidentical cones, such as the $\bar{\Gamma}$ cone and one $\bar{X}$ cone in {\sm}, we find intercone scattering to be always weak (even though peaks are in principle allowed, but are not supported by the spin structure within our model).

We believe that these considerations will be useful for the analysis of future QPI experiments on {\sm}, {\pu}, and other TI materials with multiple Dirac cones.


\bibliographystyle{apsrev4-1}
\bibliography{tki}